\PassOptionsToPackage{sort&compress}{natbib}
\documentclass[final,3p,times]{elsarticle}
\usepackage{amsmath,amssymb}
\usepackage{graphicx}
\usepackage{xcolor}
\usepackage{placeins}
\usepackage[hidelinks]{hyperref}
\journal{Journal of Subatomic Particles and Cosmology}
\graphicspath{{figures/}}
\newcommand{\etas}{\eta^0_{s\bar s}}
\hypersetup{pdfauthor={Heng-Tong Ding, Dan Zhang},
  pdftitle={Pseudoscalar meson masses in magnetic fields at zero temperature from lattice QCD}}

\begin{document}
\begin{frontmatter}
\title{Pseudoscalar meson masses in magnetic fields at zero temperature from lattice QCD}
\author[ccnu]{Heng-Tong Ding}
\author[ccnu]{Dan Zhang\fnref{poster}}
\fntext[poster]{Poster presenter}
\affiliation[ccnu]{organization={Key Laboratory of Quark and Lepton Physics (MOE) and Institute of Particle Physics, Central China Normal University},city={Wuhan},postcode={430079},country={China}}
\begin{abstract}
We study neutral and charged pseudoscalar meson masses in background magnetic fields in $(2+1)$-flavor QCD at zero temperature and physical quark masses. Using the highly improved staggered quark action at four lattice spacings, we obtain continuum-extrapolated masses from connected correlators for pions, kaons, and the fictitious $s\bar s$ pseudoscalar $\etas$. Neutral pseudoscalar masses decrease with increasing magnetic field, with the largest relative reduction for the neutral pion. In contrast, charged pion and kaon masses initially rise but tend to level off or decrease at stronger fields, departing from the point-particle lowest-Landau-level expectation. To investigate the origin of this field dependence, we compare neutral and charged pion masses at finite lattice spacings with the field included only in valence propagators or only in the sea-quark determinant. Valence-only masses approximately reproduce the full results, while sea-only masses remain close to their corresponding zero-field values.
\end{abstract}
\begin{keyword}
lattice QCD \sep magnetic fields \sep pseudoscalar mesons \sep meson masses
\end{keyword}
\end{frontmatter}

\section{Introduction}
Strong magnetic fields provide a setting for studying the nonperturbative dynamics of QCD. Such fields are relevant to off-central heavy-ion collisions~\cite{Deng:2012pc,Skokov:2009qp} and magnetars~\cite{Duncan:1992hi}. Their influence on chiral symmetry breaking and the QCD transition is illustrated by magnetic catalysis at zero temperature, inverse magnetic catalysis near the thermal crossover, and a decrease in the pseudocritical temperature with increasing magnetic field~\cite{Bali:2012zg,Bali:2011qj,DElia:2018xwo,Ding:2022tqn}.

Hadron masses provide a complementary probe of QCD dynamics in a magnetic background, with light pseudoscalar mesons closely connected to chiral symmetry breaking. A point-particle description predicts a field-independent mass for neutral pseudoscalars and an increasing lowest-Landau-level (LLL) energy for charged ones, providing a reference for examining the effects of meson compositeness~\cite{Johnson:1949wmq,Canuto:1968apg,Bali:2017ian}. Earlier $(2+1)$-flavor lattice calculations found decreasing neutral masses and nonmonotonic charged pion and kaon masses, but used a heavier-than-physical pion and a single lattice spacing~\cite{Ding:2020hxw}. Calculations at physical quark masses with continuum extrapolations are therefore needed to test these trends and establish quantitative benchmarks for effective descriptions.

These proceedings are based on our recent $(2+1)$-flavor lattice QCD study reported in Ref.~\cite{Ding:2026qzu}. We discuss continuum-extrapolated results for neutral and charged pseudoscalar meson masses in magnetic fields at zero temperature and physical quark masses.

\section{Temporal correlators and meson masses}
\label{sec:correlators}
For a pseudoscalar meson channel $H$, we define the interpolator $\mathcal M_H=\bar q_i\gamma_5q_j$, where $i,j\in\{u,d,s\}$ label quark flavors. On a fixed gauge configuration $U$, $\mathcal G_H(B;\tau;U)$ denotes the connected contraction of $\mathcal M_H(\boldsymbol x,\tau)\mathcal M_H^\dagger(\boldsymbol0,0)$, summed over the sink position $\boldsymbol x$, at Euclidean time separation $\tau$. The ensemble-averaged temporal correlator is
\begin{equation}
G_H(B;\tau)=\left\langle\mathcal G_H(B;\tau;U)\right\rangle_B .
\label{eq:correlator}
\end{equation}
The subscript $B$ denotes the gauge-ensemble average including the field-dependent fermion determinant. We retain only connected contractions in all channels, neglecting disconnected contributions. This approximation is motivated by an earlier quenched neutral-pion calculation that found the disconnected correlator consistent with zero within uncertainties~\cite{Luschevskaya:2015cko}. For the neutral pion, we extract $M_{\pi^0}$ from the flavor-averaged correlator $G_{\pi^0}=(G_{\pi_u^0}+G_{\pi_d^0})/2$, where $G_{\pi_u^0}$ and $G_{\pi_d^0}$ denote the connected $u\bar u$ and $d\bar d$ correlators, respectively~\cite{Ding:2020hxw}.

For staggered fermions, we use the corresponding local pseudoscalar operator and fit the temporal correlator with nonoscillating and opposite-parity oscillating contributions~\cite{Bazavov:2019www,Ding:2020hxw,Ding:2022tqn,Ding:2025pbu},
\begin{equation}
\begin{aligned}
G_H(n_\tau)={}&\sum_{i=1}^{N_{\rm nosc}}
A_{\rm nosc,i}\cosh\!\left[M_{\rm nosc,i}
\left(n_\tau-\tfrac{N_\tau}{2}\right)\right]
-(-1)^{n_\tau}\sum_{i=1}^{N_{\rm osc}}
A_{\rm osc,i}\cosh\!\left[M_{\rm osc,i}
\left(n_\tau-\tfrac{N_\tau}{2}\right)\right].
\end{aligned}
\label{eq:oscillatory}
\end{equation}
Here $n_\tau=\tau/a$, with $a$ the lattice spacing, and $N_\tau$ is the number of temporal lattice sites. The cosh terms include forward and backward propagation. The fit contains at least one nonoscillating state ($N_{\rm nosc}\geq1$) and zero or more oscillating states ($N_{\rm osc}\geq0$); the second sum is absent when $N_{\rm osc}=0$. The $A$ coefficients are the cosh amplitudes and the $M$ parameters are masses in lattice units. The lowest nonoscillating state gives $M_H(B)=M_{\rm nosc,1}/a$, the mass of the lowest state coupled to the pseudoscalar interpolator. We perform correlated multistate fits and select fit models and time windows with the corrected Akaike information criterion (AICc)~\cite{cavanaugh1997unifying}. We also extract masses with the oblique Lanczos method as a complementary determination~\cite{Ostmeyer:2024qgu}.

To distinguish the two ways in which the magnetic field enters a meson correlator, we introduce valence-only and sea-only calculations~\cite{DElia:2011koc,Bruckmann:2013oba}. In the full correlator, $G_H^{\rm full}\equiv G_H$, the field enters both the quark propagators and the fermion determinant governing gauge-field sampling. The two auxiliary correlators are
\begin{equation}
\begin{aligned}
G_H^{\rm val}(B;\tau)&=\langle\mathcal G_H(B;\tau;U)\rangle_0,\\
G_H^{\rm sea}(B;\tau)&=\langle\mathcal G_H(0;\tau;U)\rangle_B .
\end{aligned}
\label{eq:valence}
\end{equation}
The valence-only calculation uses finite-field propagators on zero-field ensembles, while the sea-only calculation uses zero-field propagators on finite-field ensembles. For these comparisons, all three sets of masses are obtained using only AICc-selected fits, with the same fit windows. The auxiliary masses are not additive pieces of the full mass, and their comparisons are made at finite lattice spacings.

For the full mass spectra, we extrapolate the AICc and Lanczos determinations to the continuum at each magnetic field using linear and quadratic fits in $a^2$. The four continuum estimates are averaged within each bootstrap sample. We quote the median of the resulting bootstrap distribution as the central value and take the larger of the distances from the median to the 16th and 84th percentiles as the symmetric uncertainty. Details of the fits and continuum analysis are given in Ref.~\cite{Ding:2026qzu}.

\section{Lattice setup}
\label{sec:setup}
We simulate $(2+1)$-flavor QCD using the highly improved staggered quark (HISQ) action and a tree-level improved Symanzik gauge action~\cite{Follana:2006rc,Luscher:1984xn}. The strange-quark mass $m_s$ is tuned to its physical value, and the degenerate light-quark masses satisfy $m_u=m_d\equiv m_l=m_s/27$. The four lattice spacings, $a\simeq0.112$, $0.084$, $0.067$ and $0.056$ fm, correspond to $24^3\times48$, $32^3\times64$, $40^3\times80$ and $48^3\times96$ lattices, respectively. All lattices satisfy $N_x=N_y=N_z\equiv N_\sigma$ and $N_\tau=2N_\sigma$, with an approximately fixed spatial extent $L=aN_\sigma\simeq2.7$ fm.

A uniform magnetic field along the $z$ direction is introduced by multiplying the gauge links in the fermion matrix by flavor-dependent $U(1)$ phases~\cite{DElia:2010abb,Smit:1986fn,Al-Hashimi:2008quu,Bali:2011qj}. Periodic spatial boundary conditions imply
\begin{equation}
eB=\frac{6\pi N_b}{N_xN_y a^2},\qquad N_b\in\mathbb Z,
\label{eq:flux}
\end{equation}
where $e>0$ is the elementary charge and the smallest quark charge magnitude is $e/3$. We use $N_b\in\{0,2,4,6,8,10,12\}$, covering $0\leq eB\lesssim1.22\,\mathrm{GeV}^2$. Since the spatial extent is nearly fixed, a given $N_b$ corresponds to approximately the same physical field at all four spacings.

\section{Results and discussion}
\label{sec:results}
\begin{figure}[!htbp]
\centering
\includegraphics[width=0.32\linewidth]{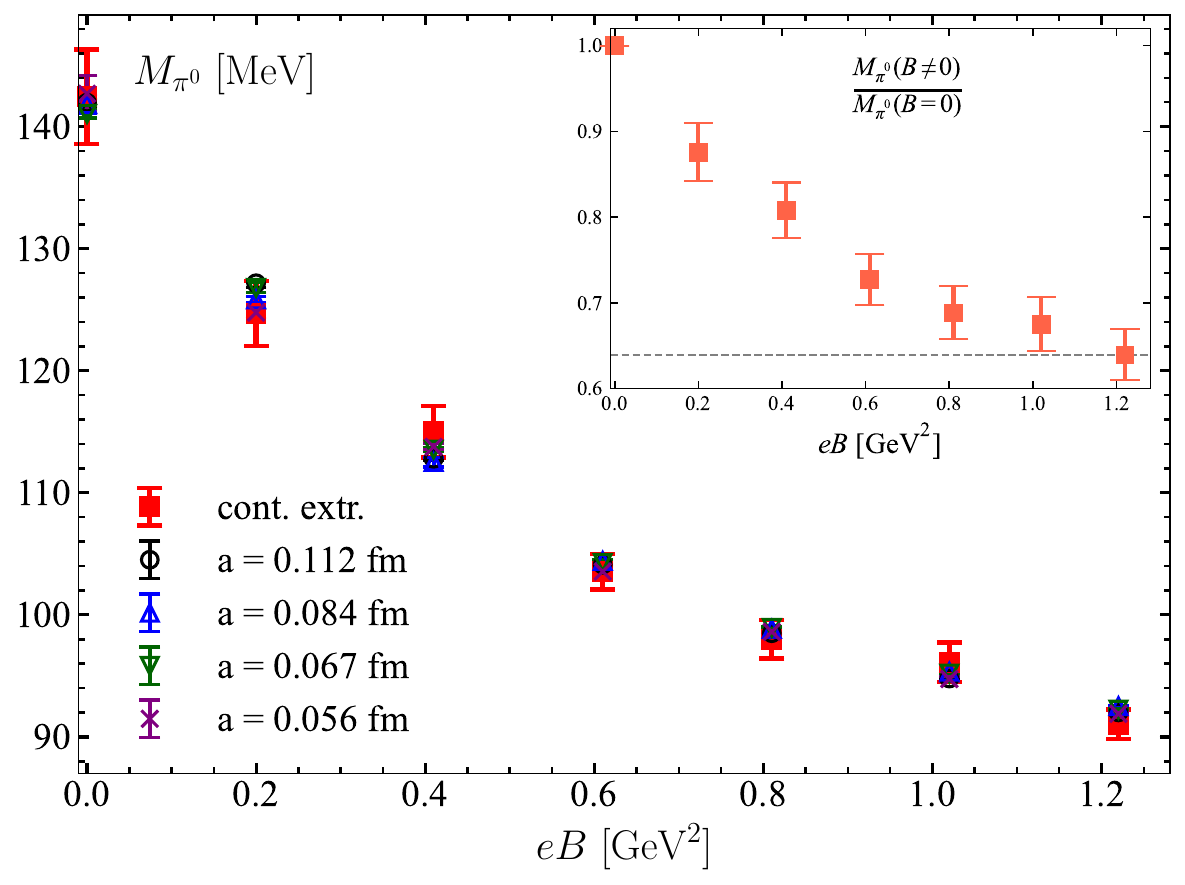}\hfill
\includegraphics[width=0.32\linewidth]{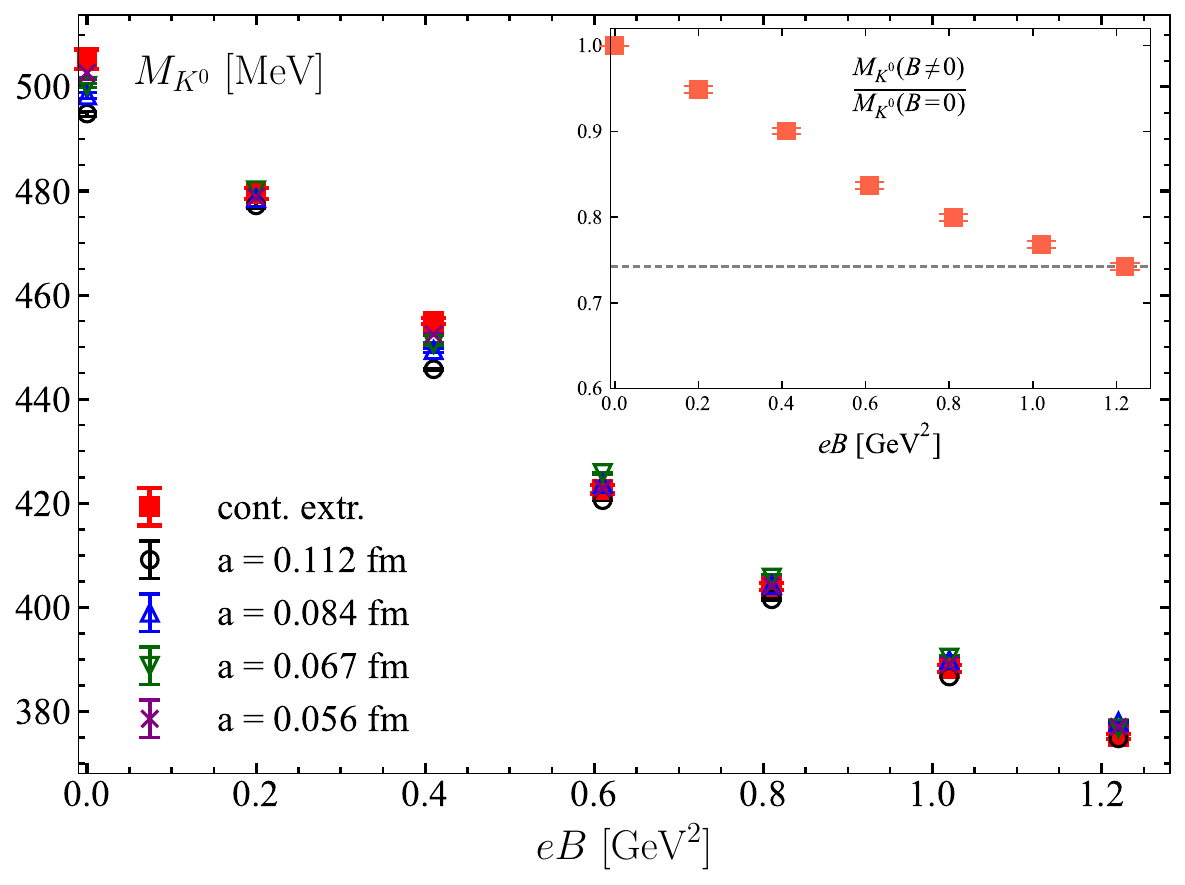}\hfill
\includegraphics[width=0.32\linewidth]{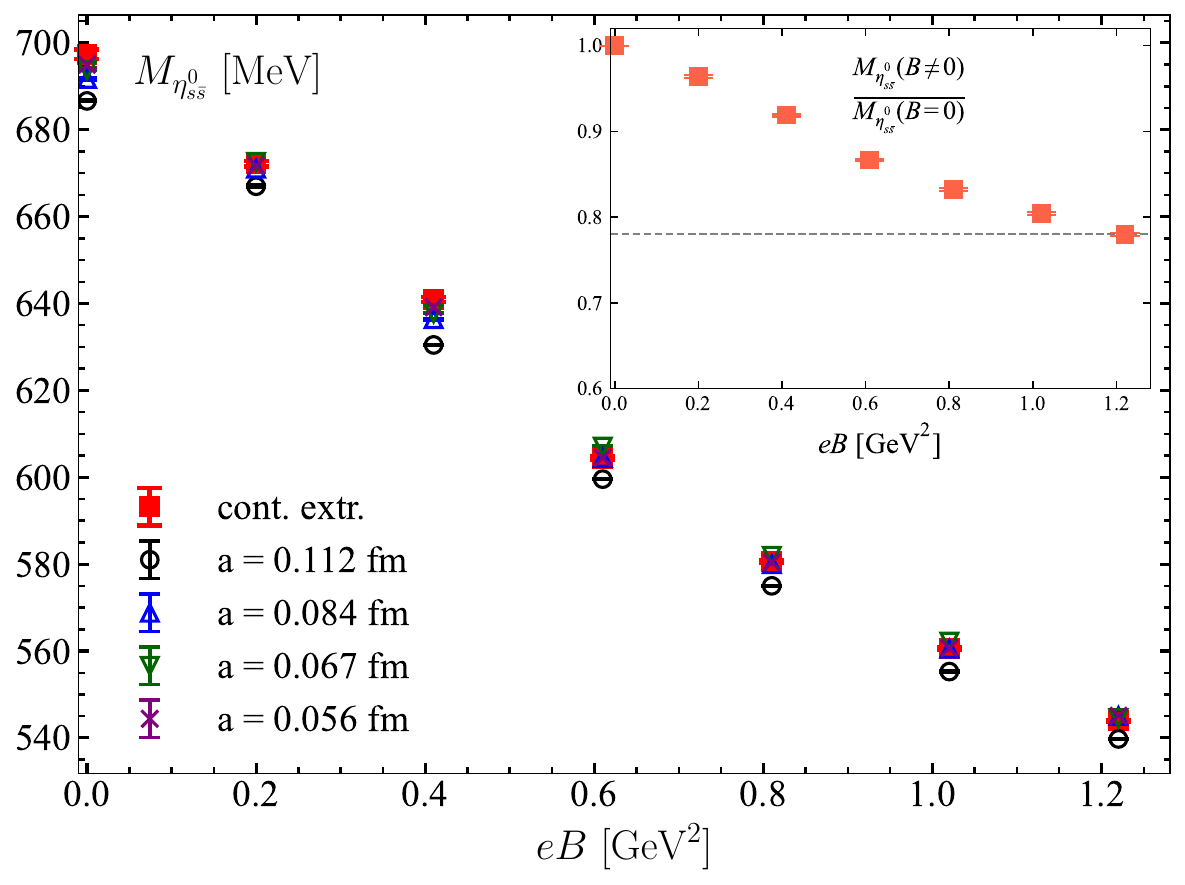}
\caption{Neutral pseudoscalar masses versus $eB$: $\pi^0$ (left), $K^0$ (middle) and $\etas$ (right). Open symbols denote finite-$a$ AICc results; filled red symbols denote combined continuum estimates. Insets give $M_H(B)/M_H(0)$. Data are from Ref.~\cite{Ding:2026qzu}.}
\label{fig:neutral}
\end{figure}

The masses of $\pi^0$, $K^0$ and $\etas$ decrease with increasing $eB$, as shown in Fig.~\ref{fig:neutral}. To compare the relative changes across channels with different zero-field masses, we consider the normalized masses $M_H(B)/M_H(0)$ in the insets. At $eB\simeq1.22\,\mathrm{GeV}^2$, the continuum-extrapolated masses are approximately $64\%$, $74\%$ and $78\%$ of their zero-field values, respectively, with the largest relative reduction in the neutral-pion channel. These reductions contrast with the field-independent masses expected for neutral pointlike pseudoscalars. Although these mesons carry no net electric charge, the magnetic field couples to their quark constituents and affects their bound-state dynamics, revealing the importance of their composite structure in magnetic fields~\cite{Avancini:2016fgq,Kojo:2026delineating}.

\begin{figure}[!htb]
  \centering
  \includegraphics[width=0.48\linewidth]{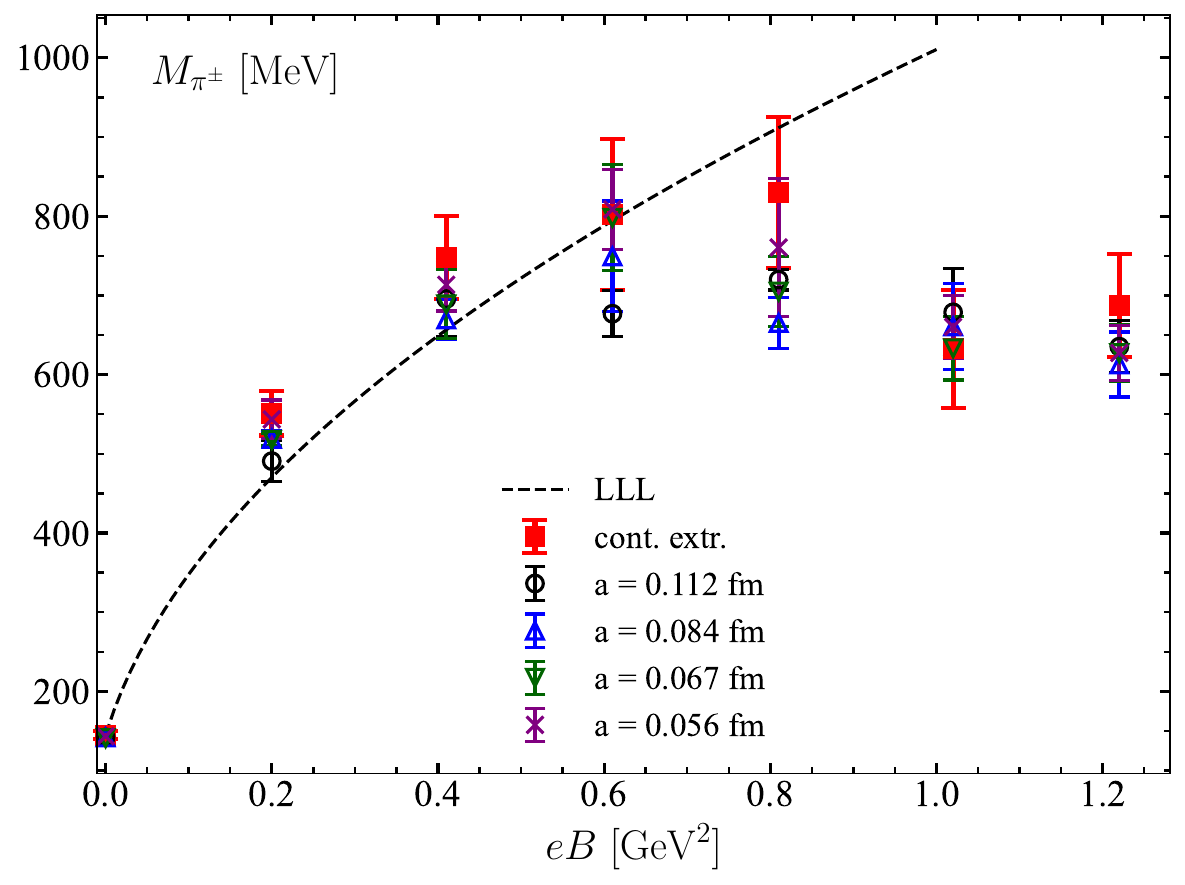}\hfill
  \includegraphics[width=0.48\linewidth]{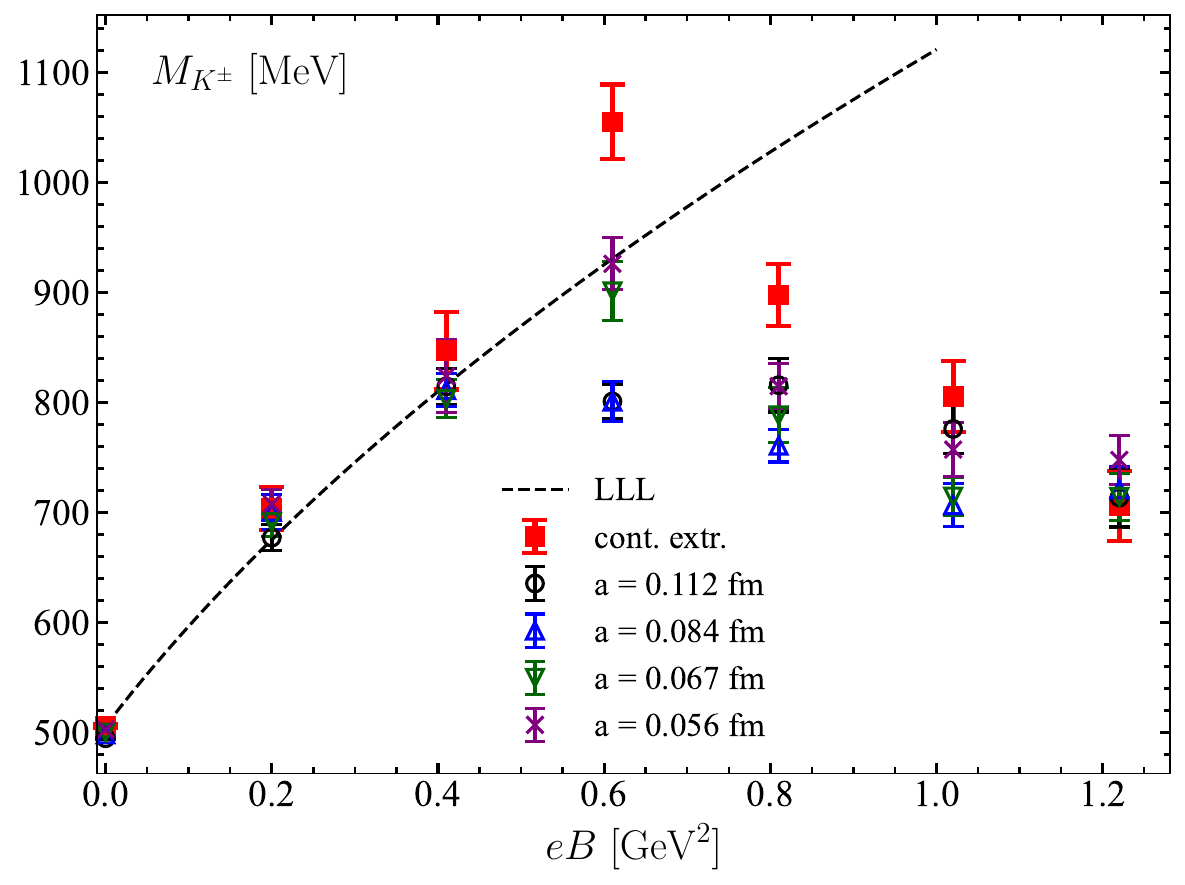}
  \caption{Charged pion (left) and kaon (right) masses versus $eB$. Symbols are as in Fig.~\ref{fig:neutral}. Dashed curves denote the point-particle LLL expectation, Eq.~\eqref{eq:lll}. Data are from Ref.~\cite{Ding:2026qzu}.}
  \label{fig:charged}
\end{figure}

The charged pion and kaon masses initially increase with $eB$, as shown in Fig.~\ref{fig:charged}. To compare this behavior with a point-particle reference, we consider the spin-zero LLL expression at vanishing longitudinal momentum~\cite{Bali:2017ian},
\begin{equation}
E_{\rm LLL}(B)=\sqrt{M_H^2(0)+|eB|},
\label{eq:lll}
\end{equation}
where $M_H(0)$ is the corresponding zero-field mass. This reference, shown by the dashed curves, increases monotonically with $eB$ and qualitatively describes the initial rise of both meson masses. At stronger fields, however, both channels depart from this monotonic rise, with masses leveling off or turning downward. The high-field decrease is clearer for the kaon, while the charged-pion results show larger uncertainties.

These results show that a point-particle description with a fixed zero-field mass cannot capture the charged spectra over the full field range.
The nonmonotonic charged spectra have motivated recent studies of meson-molecule formation in strong fields~\cite{Kojo:2026molecules} and of $\pi$--$\rho$ mixing in the charged-pion channel~\cite{Wang:2026mixing}.

Having established the different neutral and charged responses, we compare the full pion masses at each lattice spacing with those extracted from the valence-only and sea-only correlators defined in Eq.~\eqref{eq:valence}. This comparison tests the relative importance of the magnetic field acting through quark propagators and through gauge-field sampling.

\begin{figure}[!htb]
\centering
\includegraphics[width=0.48\linewidth]{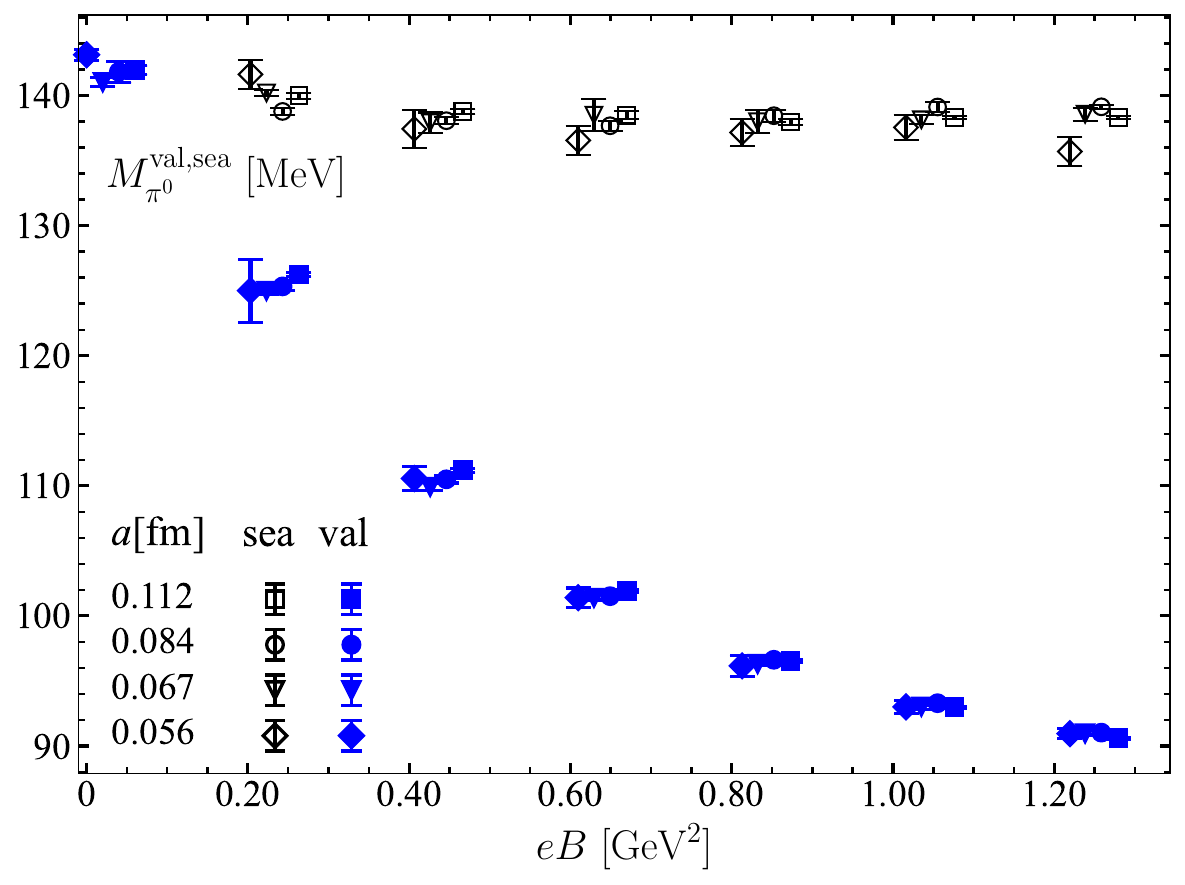}\hfill
\includegraphics[width=0.48\linewidth]{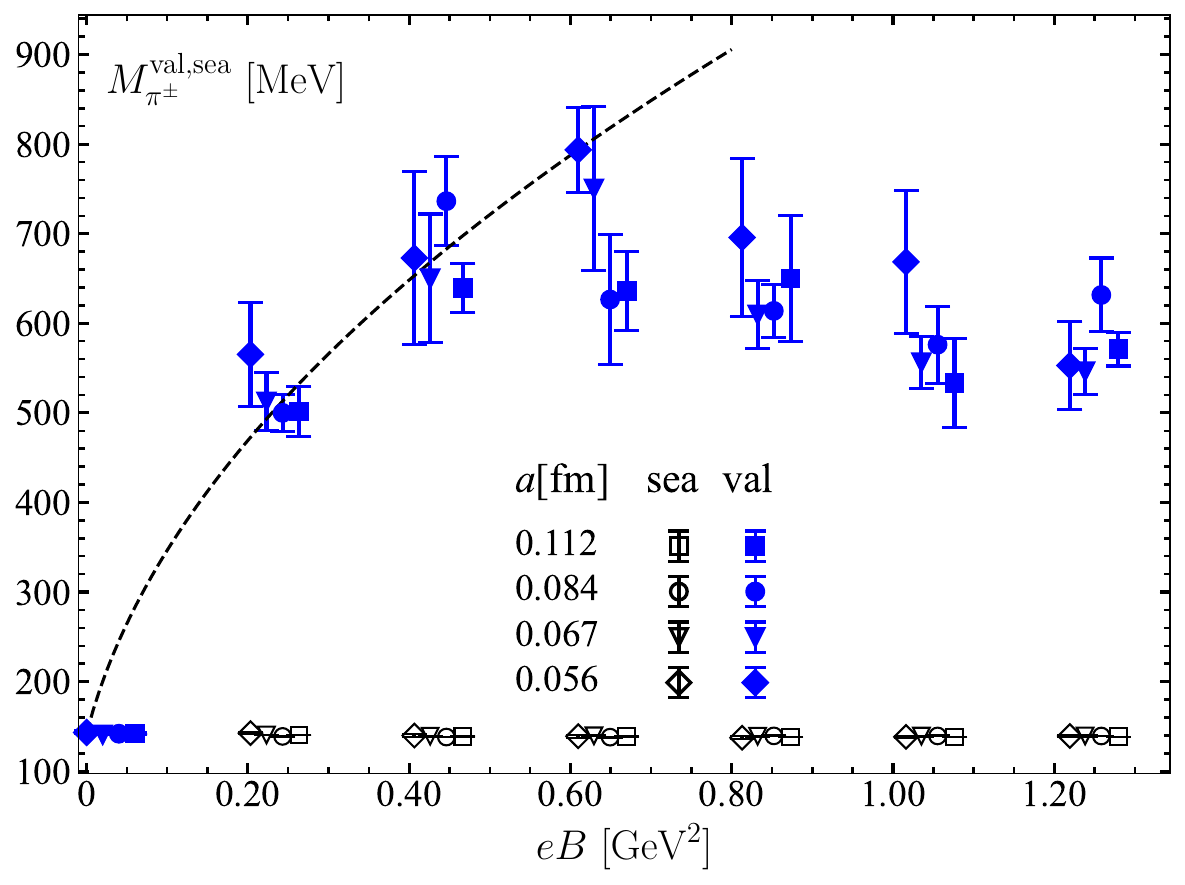}
\caption{Valence-only (filled symbols) and sea-only (open symbols) pion masses from AICc fits at four finite lattice spacings: neutral (left) and charged (right). Shapes distinguish lattice spacings; points are offset horizontally for clarity. The dashed curve is the charged point-particle LLL expectation. Data are from Ref.~\cite{Ding:2026qzu}.}
\label{fig:valence}
\end{figure}

Valence-only and sea-only masses exhibit markedly different field dependences at the four finite lattice spacings shown in Fig.~\ref{fig:valence}. Sea-only masses remain close to the corresponding zero-field masses at the same lattice spacing. Valence-only masses reproduce the neutral-pion decrease and the charged-pion nonmonotonic behavior, approximately matching the full finite-$a$ results in Figs.~\ref{fig:neutral} and~\ref{fig:charged}. This common pattern across the lattice spacings supports valence dominance of the magnetic-field dependence of neutral and charged pion masses in the zero-temperature ensembles studied here. The corresponding comparisons for neutral and charged kaons exhibit a similar pattern, as reported in Ref.~\cite{Ding:2026qzu}. This contrasts with the sea-quark-driven inverse magnetic catalysis of the light-quark condensate near the thermal crossover~\cite{Bali:2012zg,Bruckmann:2013oba,DElia:2018xwo,Ding:2022tqn}.

\section{Conclusions}
We have investigated the magnetic-field dependence of pseudoscalar meson masses using lattice QCD at zero temperature and physical quark masses. In the continuum-extrapolated spectra, neutral masses decrease with $eB$, with the largest relative reduction for the neutral pion, whereas charged pion and kaon masses initially rise and then level off or decrease at stronger fields. The charged spectra thus depart from the point-particle LLL expectation based on fixed zero-field masses, highlighting the role of internal structure and interactions.

At finite lattice spacings, valence-only masses approximately reproduce the full magnetic-field dependence of neutral and charged pion masses, while sea-only masses remain close to their zero-field values. These comparisons support valence dominance of the pion mass response in the ensembles studied here. The results provide quantitative benchmarks for effective descriptions of mesons in magnetic fields.

\section*{Acknowledgments}
We acknowledge support from the National Natural Science Foundation of China (Grants 12293064, 12293060 and 12325508), the National Key Research and Development Program of China (Contract 2022YFA1604900), and the Fundamental Research Funds for the Central Universities, Central China Normal University (Grants 30101250314 and 30106250152). Simulations were performed at the Nuclear Science Computing Center at Central China Normal University ($\mathrm{NSC}^3$) and Wuhan Supercomputing Center.

\bibliographystyle{elsarticle-num}
\bibliography{refs}
\end{document}